# Secure Direct Communication based on Ping-pong Protocol

Arti Chamoli and C. M. Bhandari
Indian Institute of Information Technology, Allahabad, Deoghat, Jhalwa,
Allahabad-211011, India.
Email: achamoli@iiita.ac.in, cmbhandari@yahoo.com

**Abstract**
Communication security with quantum key distribution has been one of the important features of quantum information theory. A novel concept of secured direct communication has been the next step forward without the need to establish any shared secret key. The present paper is based on the ping – pong protocol with three particle GHZ state as the initial state where the receiver can simultaneously receive information from two parties. Possible eavesdropping on the travelling particle has been examined.

**Introduction**

The security of the quantum cryptographic protocols is based on the establishment of a secret key shared by two legitimate remote communicators prior to the transmission of any confidential message. The key is then used for encryption and decryption of message. It cannot although prevent eavesdropping but can check any tampering on the grounds of unusual error rate. Origin of quantum key distribution in its oldest form is accredited to Bennett and Brassard [1] in 1984. This pioneering work was named as BB84 protocol. According to this protocol, one of the two users of the quantum information channel sends quantum bits to the other. The quantum bits are transmitted either in $|0\rangle, |1\rangle$ basis or the diagonal basis constituted by $\frac{1}{\sqrt{2}}(|0\rangle+|1\rangle), \frac{1}{\sqrt{2}}(|0\rangle-|1\rangle)$. The receiver then performs a measurement in either of the two bases. After discussing the results of their measurements, the users form a string of quantum bits for which the basis of measurement was same. This string serves the purpose of shared secret key for both ciphering and deciphering the secret message. Since its inception, there has been constant advancement in this area. In [2] Bennett proposed quantum cryptography using any two non-orthogonal states where the two parties share no secret initially. Ekert [3] reported a secret key generation scheme based on generalized Bell's theorem. This was further modified in [4] describing a simpler EPR scheme for quantum cryptography without invoking Bell's theorem. Generalization of BB84 quantum cryptographic protocol using three conjugate bases [5], quantum key distribution in the holevo limit [6] and extension of BB84 protocol in terms of encoding in N-dimensional Hilbert space [7] are some of the developments in the context of quantum key distribution. A detailed review of the various quantum secret sharing protocols is given in [8].

A new direction to the study of quantum cryptography was given by Biege et al [9] who proposed the novel concept of secure direct communication without establishing any shared secret key. Another advancement was made by Bostrom et al [10] by presenting a deterministic communication scheme using entanglement. Quantum direct communication protocol using blocks of EPR pairs [11], secure direct communication by EPR pairs and entanglement swapping [12], secure direct communication with a quantum one-time pad using single photons [13], deterministic direct quantum communication by using entanglement swapping and local unitary operations [14] are the major achievements in the field of quantum cryptography to date. In addition two way communication using EPR was presented by Nguyen [15] and recently a scheme based on entanglement swapping for transmission of information from many parties to a common party was presented by Gao et al [16]. Simultaneous Quantum secure direct communication among three parties was proposed by Xing-Ri Jin et al [17] using three particle GHZ states.

In this letter, a secure direct communication scheme based on ping-pong protocol has been presented for a three particle GHZ state. The eavesdropper's attack operation considered in [18] has been incorporated to affirm the advantage of a GHZ state over a two particle singlet as the initial state. To start with, we give a brief review of Bostrom and Felbinger's ping-pong protocol taking modifications as introduced in [18,19] into consideration. Bob is in possession of two photons in an entangled state $|\psi^+\rangle = 1/\sqrt{2}(|0\rangle|1\rangle + |1\rangle|0\rangle)$ of the polarization degree of freedom. He keeps one photon (the home photon) with himself and sends the other one (the travel photon) to Alice through a quantum channel. Alice randomly chooses to be either in message mode or control mode. In message mode, she can communicate 0 or 1. She does nothing i.e. performs an identity operation $I = |0\rangle\langle 0| + |1\rangle\langle 1|$ operation on the travel photon to communicate 0 to Bob. Else, she performs $\sigma_z = |0\rangle\langle 0| - |1\rangle\langle 1|$ operation on the received photon to communicate 1 to Bob. The application of $\sigma_z$ to $|\psi^+\rangle$ transforms the state to $|\psi^-\rangle = 1/\sqrt{2}(|0\rangle|1\rangle - |1\rangle|0\rangle)$. On receiving back the travel photon, Bob performs a Bell measurement on the two photons, which results in the final state as $|\psi^+\rangle$ or $|\psi^-\rangle$. Accordingly he infers the encoded message to be 0 or 1 respectively. Alternatively, if Alice is in control mode to check any possible eavesdropping she measures the polarization state of the travel photon in the z-basis $= \{|0\rangle, |1\rangle\}$. She informs Bob about her measurement result who also switches to control mode and performs a measurement in the same z-basis. In absence of any eavesdropper the photons remain in the original anti-correlated state. Thus variation of any kind indicates the presence of an eavesdropper and the transmission is immediately aborted. In [19] Wojcik pointed out that the protocol is not secure for transmission efficiencies lower than 60% by introducing an eavesdropping scheme and suggested a way to improve the original ping-pong protocol. This eavesdropping scheme was further improved in [18] by defining a new attack operation. The improved eavesdropping scheme reduced eavesdropping induced channel loss, therefore eavesdropper can attack all the transmitted bits in a larger domain of quantum channel efficiency.

**Ping-Pong Protocol with GHZ**

Let us now discuss the communication protocol starting with a three particle GHZ state as the initial state. In the present scheme Bob and Charlie who are at two different places can communicate secret messages to Alice. She can receive one bit of information from

Bob and two bits of information from Charlie simultaneously through separate quantum channels. It has been assumed that an eavesdropper is present on one of the routes at the most.

Alice prepares the initial state with three photons which can be one of the eight three-particle GHZ states. The eight GHZ states are:

$$|\psi_1\rangle = \frac{1}{\sqrt{2}}(|000\rangle_{ABC} + |111\rangle_{ABC}) \quad |\psi_5\rangle = \frac{1}{\sqrt{2}}(|010\rangle_{ABC} + |101\rangle_{ABC})$$

$$|\psi_2\rangle = \frac{1}{\sqrt{2}}(|000\rangle_{ABC} - |111\rangle_{ABC}) \quad |\psi_6\rangle = \frac{1}{\sqrt{2}}(|010\rangle_{ABC} - |101\rangle_{ABC})$$

$$|\psi_3\rangle = \frac{1}{\sqrt{2}}(|100\rangle_{ABC} + |011\rangle_{ABC}) \quad |\psi_7\rangle = \frac{1}{\sqrt{2}}(|110\rangle_{ABC} + |001\rangle_{ABC})$$

$$|\psi_4\rangle = \frac{1}{\sqrt{2}}(|100\rangle_{ABC} - |011\rangle_{ABC}) \quad |\psi_8\rangle = \frac{1}{\sqrt{2}}(|110\rangle_{ABC} - |001\rangle_{ABC})$$

(1)

Bob and Charlie can perform either of the four unitary operations:
$I = |0\rangle\langle 0| + |1\rangle\langle 1| \quad \sigma_x = |0\rangle\langle 1| + |1\rangle\langle 0|$
$i\sigma_y = |0\rangle\langle 1| - |1\rangle\langle 0| \quad \sigma_z = |0\rangle\langle 0| - |1\rangle\langle 1|$
for encoding their respective information.

The three agree on that Bob can perform unitary operations $I$ and $i\sigma_y$ in order to communicate 0 and 1 respectively to Alice. At the same time, Charlie can operate upon his qubit with unitary operations $I$, $\sigma_x$, $i\sigma_y$ and $\sigma_z$ to encode two bits of classical information as 00, 01, 10, and 11, respectively. Bob and Charlie perform their respective operations independently. Each sender can thus communicate to Alice independently. This increments the confidentiality of information to be communicated because one of the senders may be dishonest; but he will have no idea about the message of honest sender.

Suppose Alice prepares the initial state of three photons in entangled state $|\psi_5\rangle_{ABC} = \frac{1}{\sqrt{2}}(|010\rangle + |101\rangle)$. She keeps the first photon (photon A) with herself and forwards the second (photon B) and third (photon C) photons to Bob and Charlie, respectively. Bob and Charlie are at two different places so photons are transmitted through separate quantum channels. Bob and Charlie cannot make out which of the two photons, B or C, has been sent to them. Alice does not declare the order of photons to Bob and Charlie. The eavesdropper too, will not be able to mark the order of photons. This again adds to the security of communication. After the distribution of photons, Bob and Charlie mutually decide to be either in message mode and accordingly inform Alice about the mode they are in.

In the control mode, Bob and Charlie simply measure the polarization state of their respective photons in the basis $\{|0\rangle, |1\rangle\}$. After getting the measurement result they both inform it to Alice through public channel. She too changes over to control mode and measures her photon of the shared GHZ state in the same basis. Since the initial state $|\psi_5\rangle$ was known to her so if she finds that the state after measurement is the same as the initial one, she continues the communication. Otherwise, if the measurement results are not in the expected order, an eavesdropper is suspected of tampering with the photons passing through the quantum channel. Hence the communication is at once terminated.

In the message mode, Bob and Charlie are in possession of one photon each without knowing the order of their respective photons in the complete GHZ state under consideration. Going by the example with $|\psi_5\rangle_{ABC} = \frac{1}{\sqrt{2}}(|010\rangle + |101\rangle)$, Alice sends photon B to Bob and photon C to Charlie. On receiving the photons, Bob performs the operation $I$ or $i\sigma_y$ on his qubit accordingly if he has to convey 0 or 1. Charlie on his part performs either of $I$, $\sigma_x$, $i\sigma_y$ and $\sigma_z$ operations to encode 00, 01, 10 or 11 on his qubit. After performing their choiced operations, Bob and Charlie send their respective qubits back to Alice. She is now in possession of one and two bits of information from Bob and Charlie respectively. She then deciphers the precise information by performing GHZ measurements on the three qubits. Receiving any of the eight GHZ entangled states will let her know about the specific operations performed by Bob and Charlie. Table 1. shows the consequent state obtained by Alice after making a joint measurement, on her own particle and the particles sent by Bob and Charlie after performing their respective measurements, in the GHZ states expressed in (1). In the absence of an eavesdropper, using a GHZ state in place of a singlet, as in [10], enables Alice to receive two different messages simultaneously from two distant senders. The operations performed by Bob and Charlie transform the initial state into one of the eight GHZ states. Fig. 1 shows the communication protocol between Alice, Bob and Charlie in the absence of an eavesdropper.

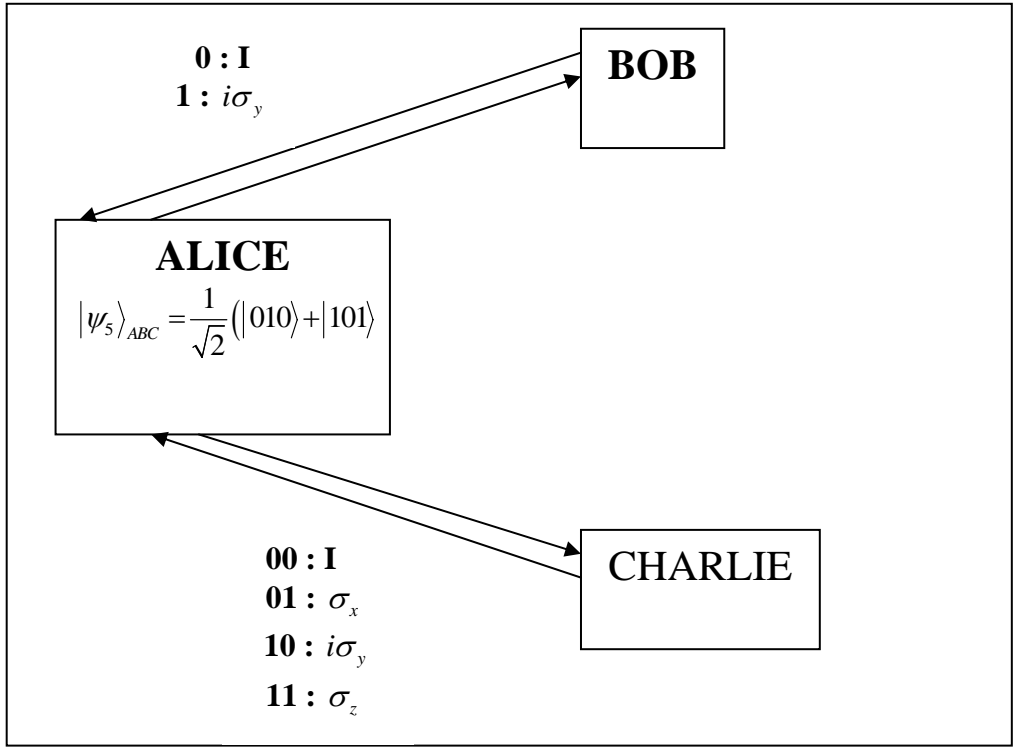

Fig. 1. Ping – pong protocol with three particle GHZ state in the absence of eavesdropper.

Table 1. Resultant GHZ state conditioned on Bob's and Charlie's measurement. The initial state is $|\psi_5\rangle_{ABC} = \frac{1}{\sqrt{2}}(|010\rangle + |101\rangle)$

| Bob's operation | Charlie's Operation | Resultant GHZ State |
|---|---|---|
| $I$ | $I$ | $\frac{1}{\sqrt{2}}(|010\rangle + |101\rangle)$ |
| $I$ | $\sigma_x$ | $\frac{1}{\sqrt{2}}(|100\rangle + |011\rangle)$ |
| $I$ | $i\sigma_y$ | $\frac{1}{\sqrt{2}}(|100\rangle - |011\rangle)$ |
| $I$ | $\sigma_z$ | $\frac{1}{\sqrt{2}}(|010\rangle - |101\rangle)$ |
| $i\sigma_y$ | $I$ | $\frac{1}{\sqrt{2}}(|000\rangle - |111\rangle)$ |
| $i\sigma_y$ | $\sigma_x$ | $(-)\frac{1}{\sqrt{2}}(|110\rangle - |001\rangle)$ |
| $i\sigma_y$ | $i\sigma_y$ | $(-)\frac{1}{\sqrt{2}}(|110\rangle + |001\rangle)$ |
| $i\sigma_y$ | $\sigma_z$ | $\frac{1}{\sqrt{2}}(|000\rangle + |111\rangle)$ |

Taking the eavesdropping probability into consideration, it can be easily figured out that the eavesdropper (say Eve) cannot approach Alice's qubit. She only has access to the traveling photons by intervening the quantum transmission from Alice to Bob and Charlie and back. In the present protocol, we have assumed that there is only one eavesdropper and since Bob and Charlie are at distant places so eavesdropping is operational in one of the quantum channels only. Following the attack operation as presented by Zhang et al [18] we examine the two scenarios where Bob's particle is captured in one and Charlie's in other by Eve.

**1. Bob's particle is captured**

The attack operation starts with the preparation of two auxiliary modes x, y where a photon in the state $|0\rangle$ is present in y mode and the x mode is kept empty thereby denoted as $|vac\rangle_x$. Hence after Eve's operation the combined state can now be written as

$$|initial\rangle = |\psi_5\rangle_{ABC} |vac\rangle_x |0\rangle_y$$

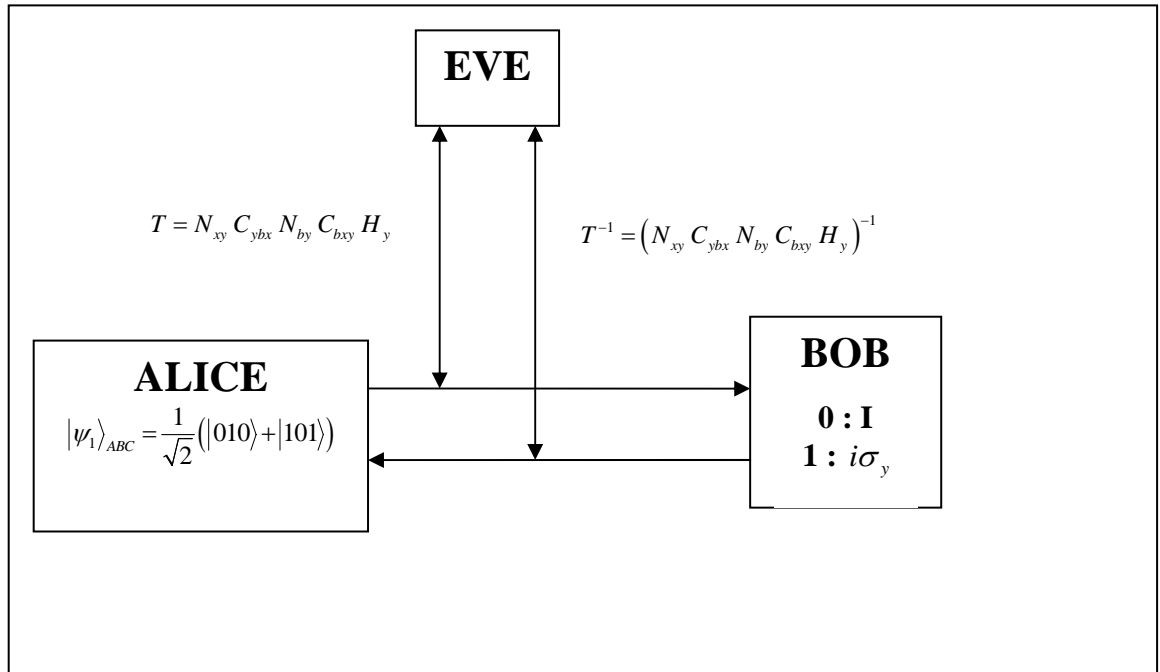

Fig. 2. Eavesdropping on Bob's particle.

The attack operation T, consists of a sequence of gates [17] as
$T = N_{xy} C_{ybx} N_{by} C_{bxy} H_y$,
where $N$ is CNOT gate, $C$ is a three mode controlled polarizing beam splitter gate, and $H$ is the hadamard gate. $C$ gate in itself consists of CNOT gates and polarizing beam splitter which transmits (reflects photons) in the state 0(1) [19]. The attack while the photon is traveling from Alice to Bob leads to the state $|A-B\rangle$, where

$$|A-B\rangle = T|initial\rangle$$
$$= \frac{1}{2}|00\rangle_{AC}\left(|vac\rangle_B|1\rangle_x|0\rangle_y + |1\rangle_B|1\rangle_x|vac\rangle_y\right)$$
$$+ \frac{1}{2}|11\rangle_{AC}\left(|0\rangle_B|vac\rangle_x|1\rangle_y + |0\rangle_B|0\rangle_x|vac\rangle_y\right)$$

(2)

The state is same as obtained by Zhang et al [18]. One can easily see that correlation of initial entangled GHZ state is not disturbed after the attack operation. Hence Eve can effectively attack without getting detected, if Alice, Bob and Charlie are in control mode. However, the attack operation increases the channel loss as Bob will find no photon with probability ¼ due to swapping of quantum channel with x mode. In the message mode, Bob performs $I$ or $i\sigma_y$ operation and sends back the photon to Alice. Meanwhile, Charlie also performs his operation on the photon received (say the $I$ operation) and sends the photon back to Alice. On the way back to Alice, Eve again attacks Bob's photon and performs $T^{-1}$ unitary operation. This attack is now referred to as B-A. The state after the second attack depends on encoding operation performed by Bob. If Bob performs $I$ operation then,

$$|B-A\rangle_0 = \frac{1}{\sqrt{2}}\left(|010\rangle_{ABC} + |101\rangle_{ABC}\right)|vac\rangle_x|0\rangle_y$$
$$= |\psi_5\rangle|vac\rangle_x|0\rangle_y$$

(3)

Otherwise, the state becomes,

$$|B-A\rangle_1 = \frac{1}{2\sqrt{2}}|010\rangle_{ABC}|vac\rangle_x\left(|0\rangle_y + |1\rangle_y\right) + \frac{1}{2}|0vac0\rangle_{ABC}|01\rangle_x|vac\rangle_y$$
$$- \frac{1}{2\sqrt{2}}|1vac1\rangle_{ABC}|1\rangle_x\left(|0\rangle_y - |1\rangle_y\right) - \frac{1}{2}|1101\rangle_{ABBC}|vac\rangle_x|vac\rangle_y$$
$$= \frac{1}{4}\left\{|\psi_5\rangle\left(|0\rangle_y + |1\rangle_y\right) + |\psi_6\rangle\left(|0\rangle_y + |1\rangle_y\right)\right\}|vac\rangle_x + \frac{1}{2}|0vac0\rangle_{ABC}|01\rangle_x|vac\rangle_y$$
$$- \frac{1}{2\sqrt{2}}|1vac1\rangle_{ABC}\left(|0\rangle_y - |1\rangle_y\right)|1\rangle_x - \frac{1}{2}|1101\rangle_{ABBC}|vac\rangle_x|vac\rangle_y$$

(4)

Equations (3) and (4) give the possible measurement results obtained by Alice and Eve. It is evident from expressions (3) and (4) that by measuring the photon in channel *y*, Eve can identify the operation performed by Bob with certainty only when he performs operation $I$ on his qubit. On the other hand, if Bob operates upon his qubit with $i\sigma_y$, Eve will be able to identify the operation performed by Bob with 1/4 probability. The inverse attack operation in the latter case results in eavesdropping –induced channel loss and quantum bit error rate (QBER). QBER, given by $(P_{01} + P_{10})$, specifies the error due to wrong transmission of information. $P_{01}$ is the probability with which the sender transmits 0 and 1 is received at the other end and $P_{10}$ is the probability with which the sender transmits 1 and 0 is received at the other end. This reduces Alice's information about both Bob's and Charlie's operation to the level of $1/4$. Fig. 2 shows the eavesdropping scheme on Bob's particle.

The eavesdropping scheme induces channel loss and Alice receives two photons instead of one through Bob's channel with a probability of $1/4$. These erroneous results in the transmission from Bob to Alice aids detection of eavesdropping. The legitimate users can set an upper bound on the channel losses due to noise. Channel loss exceeding the maximum limit, confirms the presence of an eavesdropper in case of an ideal channel.

 2. Charlie's particle is captured
Now we consider the case when Eve manipulates Charlie's photon. The initial preparations and the attack operation being same as in the previous case, gives the combined state after Eve's manipulations $|A-C\rangle$ as,

$$|A-C\rangle = \frac{1}{2}|01\rangle_{AB}\left(|0\rangle_C |vac\rangle_x |1\rangle_y + |0\rangle_C |0\rangle_x |vac\rangle_y\right) + \frac{1}{2}|10\rangle_{AB}\left(|vac\rangle_C |1\rangle_x |0\rangle_y + |1\rangle_C |1\rangle_x |vac\rangle_y\right)$$

(5)

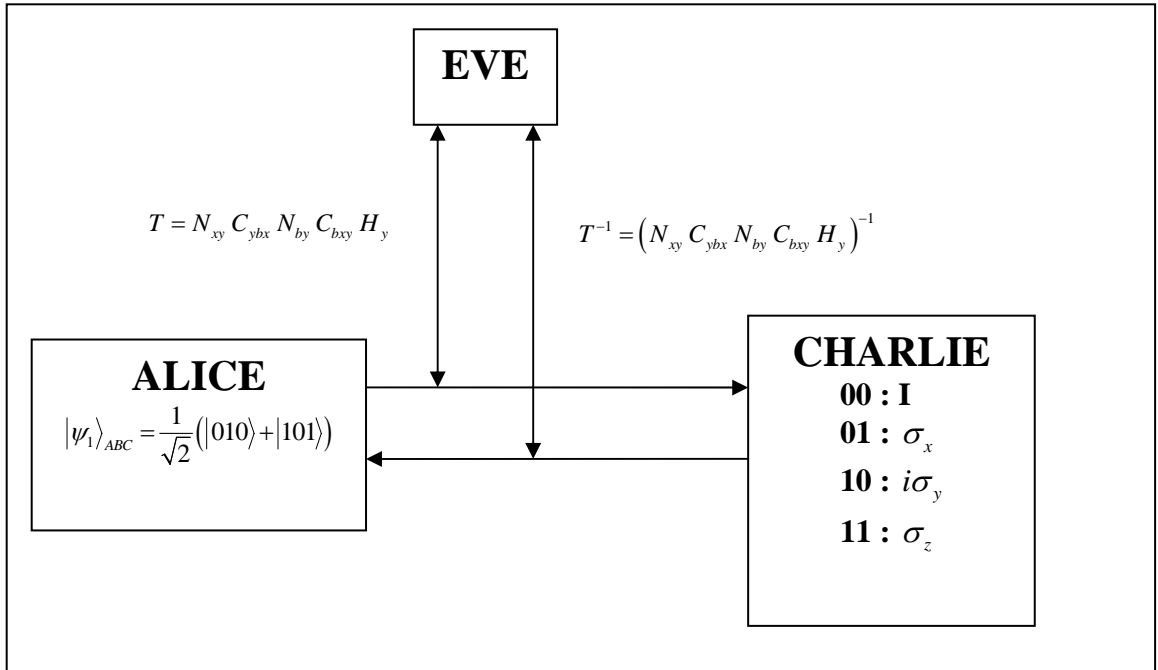

Fig. 3. Eavesdropping on Charlie's particle.

This is similar to $|A-B\rangle$, hence all the discussions made in the previous section are applicable here too. The changes however appear after the second attack operation when Charlie encodes 00, 01, 10 or 11, and sends the photon back to Alice. Encoding operations of Bob and Charlie are independent, so for the present we may assume that Bob performs unitary operation $I$ on his qubit. Writing the combined state after Charlie has encoded his information and Eve's attack as $|C-A\rangle$, we can write the four possible states depending upon Charlie's information to be 00, 01, 10, and 11 as $|C-A\rangle_{00}, |C-A\rangle_{01}, |C-A\rangle_{10}, |C-A\rangle_{11}$, respectively. The four states are,

$$|C-A\rangle_{00} = \frac{1}{\sqrt{2}}\left(|010\rangle_{ABC}|0\rangle_y + |101\rangle_{ABC}|0\rangle_y\right)|vac\rangle_x$$
$$= |\psi_5\rangle|vac\rangle_x|0\rangle_y$$

(6)

$$|C-A\rangle_{01} = \frac{1}{2\sqrt{2}}\left(|01vac\rangle_{ABC}|0\rangle_y - |01vac\rangle_{ABC}|1\rangle_y\right)|1\rangle_x + \frac{1}{2}|0110\rangle_{ABCC}|vac\rangle_x|vac\rangle_y$$
$$+ \frac{1}{2\sqrt{2}}\left(|101\rangle_{ABC}|0\rangle_y + |101\rangle_{ABC}|1\rangle_y\right)|vac\rangle_x + \frac{1}{2}|10vac\rangle_{ABC}|01\rangle_x|vac\rangle_y$$
$$= \frac{1}{2\sqrt{2}}\left(|01vac\rangle_{ABC}|0\rangle_y - |01vac\rangle_{ABC}|1\rangle_y\right)|1\rangle_x + \frac{1}{2}|0110\rangle_{ABCC}|vac\rangle_x|vac\rangle_y$$
$$+ \frac{1}{4}\left\{(|\psi_5\rangle - |\psi_6\rangle)|0\rangle_y + (|\psi_5\rangle - |\psi_6\rangle)|1\rangle_y\right\}|vac\rangle_x + \frac{1}{2}|10vac\rangle_{ABC}|01\rangle_x|vac\rangle_y$$

(7)

$$|C-A\rangle_{10} = \frac{1}{2\sqrt{2}}\left(-|01vac\rangle_{ABC}|0\rangle_y + |01vac\rangle_{ABC}|1\rangle_y\right)|1\rangle_x - \frac{1}{2}|0110\rangle_{ABCC}|vac\rangle_x|vac\rangle_y$$
$$+ \frac{1}{2\sqrt{2}}\left(|101\rangle_{ABC}|0\rangle_y + |101\rangle_{ABC}|1\rangle_y\right)|vac\rangle_x + \frac{1}{2}|10vac\rangle_{ABC}|10\rangle_x|vac\rangle_y$$
$$= \frac{1}{2\sqrt{2}}\left(-|01vac\rangle_{ABC}|0\rangle_y + |01vac\rangle_{ABC}|1\rangle_y\right)|1\rangle_x - \frac{1}{2}|0110\rangle_{ABCC}|vac\rangle_x|vac\rangle_y$$
$$+ \frac{1}{4}\left\{(|\psi_5\rangle - |\psi_6\rangle)|0\rangle_y + (|\psi_5\rangle - |\psi_6\rangle)|1\rangle_y\right\}|vac\rangle_x + \frac{1}{2}|10vac\rangle_{ABC}|10\rangle_x|vac\rangle_y$$

(8)

$$|C-A\rangle_{11} = \frac{1}{\sqrt{2}}\left(|010\rangle_{ABC}|0\rangle_y + |101\rangle_{ABC}|1\rangle_y\right)|vac\rangle_x$$
$$= \frac{1}{2}\left\{(|\psi_5\rangle + |\psi_6\rangle)|0\rangle_y + (|\psi_5\rangle - |\psi_6\rangle)|1\rangle_y\right\}|vac\rangle_x$$

(9)

Fig. 3 shows the attack operation on Charlie's particle. Eavesdropper will not gain any information through her particles in the x and y modes, because in this case Charlie sends two bits of information through single bit by applying four different unitary operations with same probability.

The attack operation however induces channel loss and a finite probability of Alice receiving two photons instead of one via Charlie's channel. These asymmetries in the transmission from Bob to Alice will lead to detection of eavesdropping. In case of an ideal channel and noisy channel too, the legitimate users can set an upper bound on the channel losses due to noise. Channel loss exceeding the maximum limit, confirms the presence of an eavesdropper.

In summary communication based on three particles GHZ state is advantageous over two qubit maximally entangled state. In absence of any eavesdropper, Alice, can simultaneously receive one and two bits of information from Bob and Charlie respectively. Although eavesdropping sabotages the communication between the legitimate users but at the same time probability of detecting eavesdropping is also enhanced. Alice can confirm the presence of an eavesdropper in the message mode too. The eavesdropping induced quantum channel loss is at the level of 1/4 when the photon is transmitted from Alice to Bob or Charlie. An additional channel loss is induced during the second attack operation on Bob's or Charlie's photon after they have performed their respective operation. These additional losses hamper the efficacy of the quantum channel which in turn points towards the presence of an eavesdropper. If the three legitimate users initially check the authenticity of their quantum channel by determining both channel loss and QBER of the quantum channel then any abnormal rise will indicate the interference by an eavesdropper. In addition, eavesdropping can also be detected by Alice when she receives two photons instead of one via Bob's and Charlie's channel with a probability of $1/8$ and $1/4$.

**Acknowledgement**

Authors are thankful to Dr. M. D. Tiwari for his keen interest and support. Arti Chamoli is thankful to IIIT, Allahbad for financial support.

**References:**

1. C. H. Bennett and G. Brassard, in Proc. IEEE Int. Conf. on Computers, systems, and signal processing, Bangalore (IEEE, New York 1984) pp. 175-179.
2. C. H. Bennett, Phys. Rev. Lett. 68, 3121 (1992)
3. A. Ekert, Phys. Rev. Lett. 67, 661 (1991)
4. C. H. Bennett, G. Brassard, N. D. Mermin, Phys. Rev. Lett. 68 (1992) 557.
5. D. Bruss, Phys. Rev. Lett. 81 (1998) 3018.
6. A. Cabello, Phys. Rev. Lett. 85 (2000) 5635.
7. M. Bourennane, A. Karlson, and G. Bjork, Phys. Rev. A 64, 012306 (2001)
8. N. Gisin, G. Ribordy, W. Tittel, and H. Zbinden, Rev. Mod. Phys. 74, 145 (2002)

9.   A. Beige, B. G. Engler, C. Kutrsiefer, H. Weinfurter, Acta Phys. Pol. A, 101 (2002) 357.
10. K. Bostrom, T. Felbinger, Phys. Rev. Lett. 89 (2002) 187902.
11. F. G. Deng, G. L. Long, X. S. Liu, Phys. Rev. A 68 (2003) 042317.
12. T. Gao, F. L. Yan, Z. X. Wang, Nuovo Cimento B 119 (2004) 313.
13. F. G. Deng, G. L. Long, Phys. Rev. A 69 (2004) 052319.
14. Z. X. Man, Z. J. Jhang, Y. Li, Chin. Phys. Lett. 22 (2005) 18.
15. B. A. Nguyen, Phys. Lett. A 328 (2004) 6.
16. T. Gao, F. L. Yan, Z. X. Wang, J. Phys. A 38 (2005) 5761.
17. Xing-Ri Jin, Xin Ji, Ying-Qiao Zhang, Shou Zhang, Suc-Kyoung Hong, Kyu-Hwan Yeon and Chung-In Um, quant-ph/0601125
18. Zhanjun Zhang, Z. Man, Yang Li, Phys. Lett. A 333 (2004) 46-50.
19. A. Wojcik, Phys. Rev. Lett. 90 (2003) 157901.